\documentclass[12pt]{article}
\usepackage{axodraw,bbold}

\catcode`@=12
\begin{document}
\thispagestyle{empty}
\begin{flushright}
UCRHEP-T430\\
April 2007\
\end{flushright}
\vspace{0.25in}
\begin{center}
{\LARGE \bf Revelations of the E$_6$/U(1)$_N$ Model:\\
Two-Loop Neutrino Mass\\ and Dark Matter\\}
\vspace{1.0in}
{\bf Ernest Ma\\}
\vspace{0.1in}
{\sl Department of Physics and Astronomy, University of California,\\ 
Riverside, California 92521, USA\\}
\vspace{0.5in}
{\bf Utpal Sarkar\\}
\vspace{0.1in}
{\sl Physical Research Laboratory, Ahmedabad 380009, India\\}
\vspace{0.75in}
\end{center}

\begin{abstract}\
The $E_6/U(1)_N$ gauge extension of the Supersymmetric Standard Model, first 
proposed by Ma, is shown to have exactly the requisite ingredients to 
realize the important new idea that dark matter is the origin of neutrino 
mass.  With the implementation of a discrete $Z_2 \times Z_2$ symmetry, 
and particle content given by three \underline{27} representations
of $E_6$, neutrino masses are naturally generated in two loops, with 
candidates of dark matter in the loops.  All particles of this 
model are expected to be at or below the TeV scale, allowing them to be 
observable at the LHC.
\end{abstract}

\newpage
\baselineskip 24pt

The $E_6/U(1)_N$ model was proposed in 1995 by one of us \cite{m96}.  It is 
a supersymmetric model with matter content given by three \underline{27} 
representations of $E_6$, and gauge interactions of the Standard Model 
plus those of $U(1)_N$, which is a linear combination of $U(1)_\psi$ and 
$U(1)_\chi$ in the decomposition:
\begin{eqnarray}
E_6 &\to& SO(10) \times U(1)_\psi, \\
SO(10) &\to& SU(5) \times U(1)_\chi.
\end{eqnarray}
In terms of the maximal subgroup $SU(3)_C \times SU(3)_L \times SU(3)_R$ of 
$E_6$, the $U(1)_N$ charge is given by \cite{m96}
\begin{equation}
Q_N = 6 Y_L + T_{3R} - 9 Y_R,
\end{equation}
where $T_{3L,3R}$ and $Y_{L,R}$ are the usual quantum numbers of the 
$SU(2) \times U(1)$ decompositions of $SU(3)_{L,R}$. The particle content 
of a \underline{27} multiplet of $E_6$ is tabulated below. 
\begin{table}[htb]
\caption{Particle content of \underline{27} of $E_6$ under $SU(3)_C \times 
SU(2)_L \times U(1)_Y$ and $U(1)_N$.}
\begin{center}
\begin{tabular}{|c|c|c|}
\hline 
Superfield & $SU(3)_C \times SU(2)_L \times U(1)_Y$ & $U(1)_N$ \\ 
\hline
$Q = (u,d)$ & (3,2,1/6) & 1 \\
$u^c$ & $(3^*,1,-2/3)$ & 1 \\ 
$e^c$ & (1,1,1) & 1 \\
\hline
$d^c$ & $(3^*,1,1/3)$ & 2 \\ 
$L = (\nu,e)$ & $(1,2,-1/2)$ & 2 \\ 
\hline
$h$ & $(3,1,-1/3)$ & $-2$ \\ 
$\bar{E} = (E^c,N^c_E)$ & $(1,2,1/2)$ & $-2$ \\ 
\hline
$h^c$ & $(3^*,1,1/3)$ & $-3$ \\ 
$E = (\nu_E,E)$ & $(1,2,-1/2)$ & $-3$ \\ 
\hline
$S$ & $(1,1,0)$ & 5 \\
\hline
$N^c$ & (1,1,0) & 0 \\ 
\hline
\end{tabular}
\end{center}
\end{table}

\newpage
There are eleven possible generic trilinear terms invariant under 
$SU(3)_C \times SU(2)_L \times U(1)_Y \times U(1)_N$.  Five are necessary 
for fermion masses, namely
\begin{equation}
Qu^c\bar{E}, ~~ Qd^cE, ~~ Le^cE, ~~ SE\bar{E}, ~~ Shh^c,
\end{equation}
for $m_u$, $m_d$, $m_e$, $m_E$, $m_h$ respectively. The other six are
\begin{equation}
LN^c\bar{E}, ~~ QLh^c, ~~ u^ce^ch, ~~ d^cN^ch, ~~ QQh, ~~ u^cd^ch^c,
\end{equation}
some of which must be absent to prevent rapid proton decay.  Hence all such 
models require an additional discrete symmetry, the simplest of which is 
of course a single $Z_2$, resulting in eight generic possibilities, as 
shown already many years ago \cite{m88}.  We consider here instead 
an exactly conserved $Z_2 \times Z_2$ symmetry as tabulated below.
\begin{table}[htb]
\caption{Particle content of \underline{27} of $E_6$ under $M$ parity  
and $N$ parity.}
\begin{center}
\begin{tabular}{|c|c|c|}
\hline 
Superfield & $M$ & $N$ \\ 
\hline
$Q,u^c,d^c$ & + & + \\
$L,e^c$ & $-$ & + \\ 
$h,h^c$ & $-$ & + \\
$E_1,\bar{E}_1,S_1$ & $+$ & + \\ 
$E_{2,3},\bar{E}_{2,3},S_{2,3}$ & $+$ & $-$ \\ 
$N^c$ & $-$ & $-$ \\ 
\hline
\end{tabular}
\end{center}
\end{table}

The resulting allowed terms corresponding to Eq.~(4) are
\begin{equation}
Qu^c\bar{E}_1, ~~ Qd^cE_1, ~~ Le^cE_1, ~~ S_1hh^c, ~~ S_1E_1\bar{E}_1, ~~ 
S_1E_{2,3}\bar{E}_{2,3}, ~~ S_{2,3}E_1\bar{E}_{2,3}, ~~ 
S_{2,3}E_{2,3}\bar{E}_1,
\end{equation}
whereas those of Eq.~(5) consist of
\begin{equation}
LN^c\bar{E}_{2,3}, ~~ QLh^c, ~~ u^ce^ch.
\end{equation}
The terms $d^cN^ch$, $QQh$, and $u^cd^ch^c$ are not allowed.  As a 
consequence, baryon number $B$ and lepton number $L$ are conserved, 
with $B=1/3$ and $L=1$ for $h$.  Proton decay is thus forbidden by exactly 
conserved $M$ parity.

Consider now the role of $N$ parity.  Without it, all three copies of 
$E$, $\bar{E}$, and $S$ are Higgs superfields, and $N^c$ are the usual 
singlet neutrinos, with Dirac masses coming from the $LN^c\bar{E}$ terms. 
Furthermore, since $N^c$ is trivial under $U(1)_N$, it is allowed to have 
a very large Majorana mass \cite{m96,ns86} as in the Standard Model.  Thus the 
observed neutrino masses are Majorana and naturally small by virtue of 
the canonical seesaw mechanism.  This model is very interesting in its 
own right, and has been explored in some detail \cite{km96,hmrs01,kmn06}.

With exactly conserved $N$ parity, there are two important new interrelated 
consequences. (I) Since $\bar{E}_{2,3}$ do not have vacuum expectation values 
(otherwise $N$ parity would be broken), there are no Dirac masses linking 
$\nu$ with $N^c$.  Hence neutrinos are massless at tree level in this 
model. (II) The lightest particle odd under $N$ is absolutely stable and 
may be considered a candidate for the dark matter of the Universe 
\cite{bhs05}.  It may also interact with $\nu$ and $N^c$ to induce a 
Majorana mass for $\nu$ in a one-loop radiative version of the seesaw 
mechanism, as first proposed by one of us \cite{m06-1}.  Variants 
of this basic idea have also been discussed \cite{knt03,kms06,m06-2,m06-3,
ks06,hkmr07}.

The idea that the Standard Model may be extended to include a second $dark$ 
scalar doublet $\Phi_2 = [H^\pm, (H^0+iA^0)/\sqrt{2}]$ (which is odd under 
some new unbroken $Z_2$ symmetry) was considered many years ago 
\cite{mpt77,dm78}.  Either $H^0$ or $A^0$ is then absolutely stable, and 
presumably an acceptable candidate for dark matter.  This simple observation 
lay dormant for almost thirty years until recently when it was revived in 
Ref.~\cite{m06-1}; then it was studied seriously for the first time in 
Ref.~\cite{bhr06} and is now receiving much wider attention 
\cite{m06-2,mg06,hnot07,ss07,glbe07,lw07}.

First it should be pointed out that $H^0$ and $A^0$ cannot be degenerate 
in mass, otherwise they interact with the $Z$ boson in the same way as a 
scalar neutrino in supersymmetry, and it has been established for a long 
time that the latter cannot be the sole source of dark matter because its 
elastic scattering cross section with nuclei is too big to satisfy present 
constraints from direct search experiments \cite{bhs05}.  In the Standard 
Model, the mass splitting of $H^0$ and $A^0$ comes from the term
\begin{equation}
{1 \over 2} \lambda_5 (\Phi_1^\dagger \Phi_2)^2 + H.c.
\end{equation}
which is also necessary for inducing a Majorana neutrino mass in one loop, 
as explained in Ref.~\cite{m06-1}.  This quartic scalar term is not allowed 
in supersymmetry, but may be obtained in one loop as the supersymmetry is 
broken softly, as shown below.

Using Eq.~(8) and the notation $\Phi_1 = (\phi^+,\phi^0)$, $\Phi_2 = 
(\eta^+,\eta^0)$, it was shown in Ref.~\cite{m06-1} that a radiative 
Majorana neutrino mass is obtained in one loop, as shown in Fig.~1.

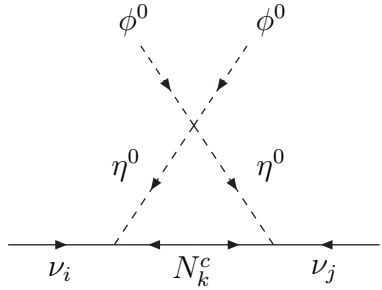
\begin{figure}[htb]
\begin{center}
\begin{picture}(360,120)(0,0)
\ArrowLine(110,10)(150,10)
\ArrowLine(180,10)(150,10)
\ArrowLine(180,10)(210,10)
\ArrowLine(250,10)(210,10)
\DashArrowLine(180,55)(150,10)3
\DashArrowLine(180,55)(210,10)3
\DashArrowLine(160,85)(180,55)3
\DashArrowLine(200,85)(180,55)3

\Text(130,0)[]{$\nu_i$}
\Text(230,0)[]{$\nu_j$}
\Text(180,0)[]{$N^c_k$}
\Text(155,40)[]{$\eta^0$}
\Text(210,40)[]{$\eta^0$}
\Text(158,95)[]{$\phi^{0}$}
\Text(210,95)[]{$\phi^{0}$}

\end{picture}
\end{center}
\caption{Would-be one-loop generation of neutrino mass.}
\end{figure}

Let $\eta^0 = (H^0+iA^0)/\sqrt{2}$, $\langle \phi^0 \rangle = v$, $m_0 = 
m(H^0)$, $m_A = m(A^0)$, $m_k = m(N^c_k)$, and assuming that $m_A^2-m_0^2 
= -2\lambda_5 v^2$ is much smaller in magnitude than $m_0^2$, the radiative 
neutrino mass matrix is given by \cite{m06-1}
\begin{equation}
({\cal M}_\nu)_{ij} = {\lambda_5 v^2 \over 8 \pi^2} \sum_k {f_{ik} f_{jk} m_k 
\over m_k^2-m_0^2} \left[ 1 - {m_k^2 \over m_k^2-m_0^2} \ln {m_k^2 \over 
m_0^2} \right],
\end{equation}
where $f_{ik}$ are the $\nu_i N^c_k \eta^0$ Yukawa couplings.

In the $E_6/U(1)_N$ model, $\lambda_5=0$ because of supersymmetry.  As the 
latter is softly broken, an effective nonzero $\lambda_5$ for the quartic 
scalar coupling $[(\tilde{N}^c_E)_1^\dagger (\tilde{N}^c_E)_{2,3}]^2$ may be 
obtained at tree level \cite{m06-3}, but here it appears only in one loop, as 
shown in Fig.~2 and Fig.~3.

\begin{figure}[htb]
\begin{center}
\begin{picture}(360,160)(0,0)
\ArrowLine(150,40)(210,40)
\ArrowLine(210,40)(210,100)
\ArrowLine(210,100)(150,100)
\ArrowLine(150,100)(150,40)
\DashArrowLine(150,40)(120,10)3
\DashArrowLine(120,130)(150,100)3
\DashArrowLine(210,100)(240,130)3
\DashArrowLine(240,10)(210,40)3

\Text(180,30)[]{$({\nu}_E)_1$}
\Text(180,110)[]{$({\nu}_E)_{2,3}$}
\Text(135,70)[]{${S}_{2,3}$}
\Text(225,70)[]{${S}_{1}$}
\Text(120,0)[]{$(\tilde{N}^c_E)_{2,3}$}
\Text(240,0)[]{$(\tilde{N}^c_E)_{1}$}
\Text(120,140)[]{$(\tilde{N}^c_E)_{1}$}
\Text(240,140)[]{$(\tilde{N}^c_E)_{2,3}$}

\end{picture}
\end{center}
\caption{One of four diagrams contributing to $\lambda_5$ from a fermion loop.}
\end{figure}
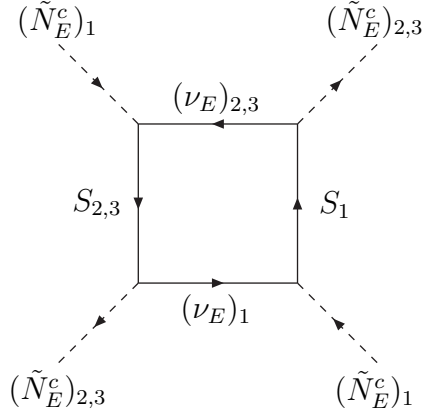

\begin{figure}[htb]
\begin{center}
\begin{picture}(360,100)(0,0)

\DashArrowArc(180,40)(30,0,180)3
\DashArrowArc(180,40)(30,180,0)3

\DashArrowLine(150,40)(120,10)3
\DashArrowLine(120,70)(150,40)3
\DashArrowLine(210,40)(240,70)3
\DashArrowLine(240,10)(210,40)3

\Text(180,0)[]{$\tilde{S}_{2,3}$}
\Text(180,80)[]{$\tilde{S}_1$}
\Text(120,0)[]{$(\tilde{N}^c_E)_{2,3}$}
\Text(240,0)[]{$(\tilde{N}^c_E)_{1}$}
\Text(120,80)[]{$(\tilde{N}^c_E)_{1}$}
\Text(240,80)[]{$(\tilde{N}^c_E)_{2,3}$}

\end{picture}
\end{center}
\caption{One of eight diagrams contributing to $\lambda_5$ from a scalar loop.}
\end{figure}
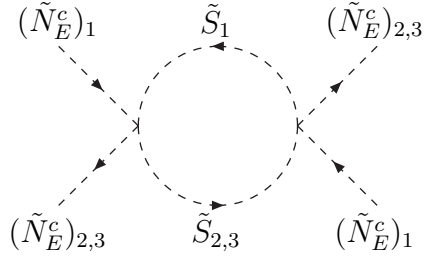

In unbroken supersymmetry, the sum of all these diagrams is exactly zero. 
As the supersymmetry is broken, an effective nonzero $\lambda_5$ will be 
obtained, and since it comes from one loop, the resulting Majorana neutrino 
mass is a two-loop effect.  This allows for the possibility of a much lower 
mass scale for $N^c$ and all the other particles appearing in the loop.  
This scenario is thus potentially verifiable at the LHC (Large Hadron 
Collider). 

The existence of $M$ parity implies the exact conservation of the usual 
$R$ parity of the MSSM (Minimal Supersymmetric Standard Model), i.e. quarks 
and leptons have even $R$, their scalar partners have odd $R$, whereas the 
scalar particles corresponding to $E$, $\bar{E}$, and $S$ have even $R$, their 
fermionic partners have odd $R$.  The neutralino sector is now a $6 \times 6$ 
mass matrix spanning the gauginos of $U(1)_N$, $U(1)_Y$, and the third 
component of $SU(2)_L$, as well as the Higgsinos $(\nu_E)_1$, $(N^c_E)_1$, 
and $S_1$.  Details have already been given in Ref.~\cite{km96}.  Without 
$N$ parity, the lightest mass eigenstate of this sector is the sole candidate 
for the dark matter of the Universe, but the situation here is more 
complicated. 

The existence of $N$ parity requires at least one more particle which 
is absolutely stable.  Consider the two classes of particles: $(R,N)=(+,-)$ 
and $(R,N)=(-,-)$.  Take the lightest one from each and consider their 
interaction with the lightest particle of the class $(R,N)=(-,+)$.  Two of 
the three must then be stable and they may or may not include the one 
associated with the MSSM. If no particle has a mass greater than the sum of 
the other two, then all three are dark-matter candidates, as pointed out 
already in Ref.~\cite{m06-3}.

The three $N^c$ fermions have $(R,N)=(+,-)$ and their scalar partners have 
$(R,N)=(-,-)$.  The six $(\nu_E)_{2,3}$, $(N^c_E)_{2,3}$, and $S_{2,3}$ 
fermions have $(R,N)=(-,-)$ and their scalar partners have $(R,N)=(+,-)$. 
In general, we expect all their masses to be of order the supersymmetry 
breaking scale.  However, there is one exception.  Consider the $6 \times 6$ 
mass matrix spanning $(\nu_E)_{2,3}$, $(N^c_E)_{2,3}$, and $S_{2,3}$.  It 
is of the form
\begin{equation}
{\cal M} = \pmatrix{0 & A & B \cr A & 0 & C \cr B & C & 0},
\end{equation}
where each entry is a $2 \times 2$ matrix.  Now $A$ comes from $\langle 
\tilde{S}_1 \rangle$, $B$ from $\langle (\tilde{N}^c_E)_1 \rangle$, and 
$C$ from $\langle (\tilde{\nu}_E)_1 \rangle$. Since $\langle \tilde{S}_1 
\rangle$ breaks $U(1)_N$, it is expected to be an order of magnitude 
greater than the others which break $SU(2)_L \times U(1)_Y$.  Hence the 
lightest particle in this sector is likely to be mostly $S_{2,3}$ with 
a mass of order $|2BC/A|$ which could be much less than 100 GeV.

The $E_6/U(1)_N$ model is defined as the linear combination of $U(1)_\psi$ 
and $U(1)_\chi$ in Eqs.~(1) and (2), under which $N^c$ is trivial.  As a 
result, the unwelcomed bilinear terms $hd^c$ and $L\bar{E}$ are admitted 
as well.  However the former is forbidden by $M$ parity and the latter 
by $N$ parity.  At the TeV energy scale, this model predicts a new $Z_N$ 
gauge boson, leptoquark scalars $\tilde{h}$ which decay into $ue$ and 
$d\nu$, singlet neutral fermions $N^c$ which decay into leptons plus 
dark-matter scalars, etc.  A rich tapestry of particles and their 
interactions awaits at the LHC.

Other variants of this model are easily perceived.  For example, if $h,h^c$ 
are even under $M$ parity, then $QLh^c$ and $u^ce^ch$ are forbidden, 
but $QQh$ and $u^cd^ch^c$ are allowed, so that $\tilde{h}^c$ become diquark 
scalars which decay into $ud$.  As for neutrino masses, the crucial term 
is $LN^c\bar{E}_{2,3}$ of Eq.~(7).  This means that even with one $N^c$, 
two nonzero masses may be obtained, which is sufficient to account for 
present neutrino-oscillation data.  The other two $N^c$ could then be 
chosen to have very small couplings so that their decay into leptons 
and antileptons would be out of thermal equilibrium and would generate 
a lepton asymmetry, which gets converted during the electroweak transition 
by sphalerons into the observed baryon asymmetry of the Universe.  Details 
will be presented elsewhere.

This work was supported in part by the U.~S.~Department of Energy under
Grant No. DE-FG03-94ER40837. US thanks the Department of Physics and 
Astronomy, University of California, Riverside for hospitality during 
a recent visit. 

\newpage

\bibliographystyle{unsrt}

\begin{thebibliography}{99}

\bibitem{m96} E. Ma, Phys. lett. {\bf B 380}, 286 (1996).

\bibitem{m88} E. Ma, Phys. Rev. Lett. {\bf 60}, 1363 (1988).

\bibitem{ns86} S. Nandi and U. Sarkar, Phys. Rev. Lett. {\bf 56}, 564 (1986).

\bibitem{km96} E. Keith and E. Ma, Phys. Rev. {\bf D54}, 3587 (1996).

\bibitem{hmrs01} T. Hambye, E. Ma, M. Raidal, and U. Sarkar, Phys. Lett. 
{\bf B512}, 373 (2001).

\bibitem{kmn06} S. F. King, S. Moretti, and R. Nevzorov, Phys. Rev. {\bf D73}, 
035009 (2006); Phys. Lett. {\bf B634}, 278 (2006).

\bibitem{bhs05} For a review, see G. Bertone, D. Hooper, and J. Silk, Phys. 
Rept. {\bf 405}, 279 (2005).

\bibitem{m06-1} E. Ma, Phys. Rev. {\bf D73}, 077301 (2006).

\bibitem{knt03} L. M. Krauss, S. Nasri, and M. Trodden, Phys. Rev. {\bf D67}, 
085002 (2003); K. Cheung and O. Seto, Phys. Rev. {\bf D69}, 113009 (2004).

\bibitem{kms06} J. Kubo, E. Ma, and D. Suematsu, Phys. Lett. {\bf B642}, 18 
(2006).

\bibitem{m06-2} E. Ma, Mod. Phys. Lett. {\bf A21}, 1777 (2006).

\bibitem{m06-3} E. Ma, Annales de la Fondation de Broglie {\bf 31}, 285 (2006) 
[hep-ph/0607142].

\bibitem{ks06} J. Kubo and D. Suematsu, Phys. Lett. {\bf B643}, 336 (2006); 
Y. Kajiyama, J. Kubo, and H. Okada, Phys. Rev. {\bf D75}, 033001 (2007).

\bibitem{hkmr07} T. Hambye, K. Kannike, E. Ma, and M. Raidal, hep-ph/0609228 
[Phys. Rev. {\bf D75}, in press]. 


\bibitem{mpt77} E. Ma, S. Pakvasa, and S. F. Tuan, Phys. Rev. {\bf D16}, 1568 
(1977).

\bibitem{dm78} N. G. Deshpande and E. Ma, Phys. Rev. {\bf D18}, 2574 (1978).

\bibitem{bhr06} R. Barbieri, L. Hall, and V. S. Rychkov, Phys. Rev. 
{\bf D74}, 015007 (2006).

\bibitem{mg06} D. Majumdar and A. Ghosal, hep-ph/0607067.

\bibitem{hnot07} L. L. Honorez, E. Nezri, J. F. Oliver, and M. H. G. Tytgat, 
JCAP {\bf 02}, 028 (2007).

\bibitem{ss07} N. Sahu and U. Sarkar, hep-ph/0701062.

\bibitem{glbe07} M. Gustafsson, E. Lundstrom, L. Bergstrom, and J. Edsjo, 
astro-ph/0703512.

\bibitem{lw07} M. Lisanti and J. G. Wacker, arXiv:0704.2816. 

\end{thebibliography}

\end{document}